\documentclass[pra,twocolumn,showpacs,superscriptaddress,floatfix]{revtex4-2}
\usepackage{graphicx}
\usepackage{epstopdf} 
\usepackage{amsmath}
\usepackage{epsfig}
\usepackage{latexsym}
\usepackage{amsfonts}
\usepackage{graphics}
\usepackage{bm}
\usepackage{braket}
\usepackage{dsfont}
\usepackage{soul}
\usepackage{ulem}
\usepackage{bbm}

\usepackage{mathtools}

\usepackage{helvet}

\graphicspath{ {./Graphics/} }
\begin{document}

\date{\today}
\author{J. Mumford}
\affiliation{Department of Physics and Astronomy, McMaster University, 1280 Main Street West, Hamilton, Ontario, Canada L8S 4M1}

\title{Many topological regions on the Bloch sphere of the spin-1/2 double kicked top}

\begin{abstract}
Floquet topological systems have been shown to exhibit features not commonly found in conventional topological systems such as topological phases characterized by arbitrarily large winding numbers.  This is clearly highlighted in the quantum double kicked rotor coupled to spin-1/2 degrees of freedom [Phys. Rev. A \textbf{97}, 063603 (2018)] where large winding numbers are achieved by tuning the kick strengths.  Here, we extend the results to the spin-1/2 quantum double kicked top and find not only does the system exhibit topological regions with large winding numbers, but a large number of them are needed to fully characterize the topology of the Bloch sphere of the top for general kick strengths.  Due to the geometry of the Bloch sphere it is partitioned into regions with different topology and the boundaries separating them are home to $0$ and $\pi$ quasienergy bound states.  We characterize the regions by comparing local versions of the mean field, quantum and mean chiral displacement winding numbers.  We also use a probe state to locate the boundaries by observing localization as the state evolves when it has a large initial overlap with bound states.  Finally, we briefly discuss the connections between the spin-1/2 quantum double kicked top and multi-step quantum walks, putting the system in the context of some current experiments in the exploration of topological phases.
\end{abstract}

\maketitle

\section{\label{Sec:Intro}Introduction}

Periodically driven systems, also called Floquet systems, have been a useful tool in simulating novel phases of matter.  Periodic driving allows for control in the time domain which can lead to exotic phases such as Anderson localization in time \cite{sacha15,giergiel17} and phase space crystals \cite{guo13,guo16}.  Of particular interest is the creation of effective magnetic fields and spin-orbit couplings with lasers to simulate topological features found in condensed matter systems \cite{goldman14a,goldman14b,eckardt15,bukov15}.  A notable example along these lines are Floquet topological insulators which are laser-induced topological states in normal materials resulting in the creation and control of chiral edge states \cite{lindner11}.  In some cases, the crystal structure of solids is also simulated using optical lattices \cite{jaksch05,dalibard11,stuhl15,creffield16}.  In these systems, the lasers creating the lattice are periodically driven to induce effective magnetic field strengths unobtainable in real materials.  This has lead to the observation of the celebrated Harper-Hofstadter model \cite{aidelsburger13,miyake13} which displays one of the more striking examples of integer quantum Hall physics.

A special class of Floquet systems involves periodic kicks rather than continuous driving.  Their appeal comes from the fact that only part of the Hamiltonian is responsible for evolving the system at a given time, so they are generally easier to conceptualize than their non-kicked counterparts.  Early applications of this method were used to explore connections between topology and chaos in the kicked Harper model \cite{leboeuf90}.  In recent years, one of the main focuses has been in controlling topological phases \cite{jangjan22}, particularly in generating large topological invariants such as the winding number which counts the number of symmetry protected states in the system \cite{cedzich18,cedzich21}. Double kicked systems such as the quantum double kicked rotor (QDKR) and the quantum double kicked top (QDKT) have shown to be promising in this regard \cite{wang08,wang09,ho12} where large topological invariants were predicted.  


Proposals involving the QDKR coupled to spin-1/2 degrees of freedom have shown that arbitrarily large winding numbers can be achieved \cite{zhou18, bolik22}.  The additional spin-1/2 degrees of freedom are significant because they put the QDKR in the perspective of quantum walks where it plays the role of a coin which is 'tossed' each step.  The state of the spin determines the direction the rotor evolves in and is analogous to tossing a coin at regular intervals and choosing a direction based on whether it is heads or tails in classical random walks \cite{aharonov93}.  Of course, the difference in the quantum case is that the coin can be in a superposition of heads and tails.  Quantum walks play an important part in the building of efficient quantum algorithms \cite {farhi98,shenvi03} and provide a foundation for quantum computation \cite{childs09}.  Quantum walks have also been proposed as a method to explore topological phases \cite{kitagawa10,groh16,sajid19} where experiments involved the measurement of the response of topological invariants to disorder \cite{cardano17,xu18} and the identification of topologically protected bound states \cite{kitagawa12,wang18,chen18}.


In this paper, we perform the natural step of extending the results of the spin-1/2 QDKR to the spin-1/2 QDKT.  This is a generalization which amounts to extending the Hilbert space of the system from a ring to a sphere \cite{haake88} - commonly referred to as the Bloch sphere.  Our main result is that, due to the change in geometry of the Hilbert space, instead of a single topological region on the ring, an arbitrarily large number of different topological regions can be generated on the Bloch sphere depending on the kick strengths.   In general, many winding numbers are needed to characterize the topology of the entire Bloch sphere and due to the bulk-boundary correspondence, the boundaries between the regions are home to protected bound states.  The number of bound states at a boundary can also be arbitrarily large, so there can be regions around the boundaries which are especially dynamically stable.  

We also show that the spin-1/2 QDKT dynamics can be framed in terms of a quantum walk on the Bloch sphere by breaking down the Floquet operator into a series of coin tosses and coin-dependent rotations.  This allows us to make proper connections between the spin-1/2 QDKT and traditional quantum walk systems as well as highlight the key difference between them which is the source of the topological boundaries.  For instance, quantum walks are often implemented with a particle in an optical lattice which has discrete translational symmetry and the boundary between two different topological regions is inserted artificially as a local breaking of that symmetry  \cite{kitagawa12,groh16,nitsche19}.  However, the Bloch sphere, does not have translational symmetry due to its geometry and we show that it is this inhomogeneity that is responsible for the topological boundaries.  Therefore, the topological boundaries of the spin-1/2 QDKT can be considered to occur naturally due to the geometry of the Bloch sphere. 

We do not choose a specific model for the top, however, some possibilities for its physical origin are the angular momentum of a single particle or Fock states of a collection of two-mode indistinguishable particles \cite{biederharn81}.  In the many-body case, the two modes can be external states like the energy levels of a trapping potential such as the system of a Bose-Einstein condensate (BEC) occupying the ground states of a double well potential.  The two modes can also be internal states like the hyperfine states of atoms \cite{zibold10,gerving12} or two polarizations of light \cite{bao20}.   In either case, the spin-1/2 degrees of freedom can represent a two-mode particle that is distinguishable from the others and is either a different species of particle with access to the same two modes \cite{mulansky11,mumford14} or the same species with a different pair of modes.  When the top operators belong to the angular momentum of a single particle, the spin-1/2 degrees of freedom can represent the internal spin states of that particle.

\section{\label{model}Model}

The system we will be investigating is the spin-1/2 QDKT which is described by the Hamiltonian

\begin{eqnarray}
\hat{H}_T &=& \frac{\Lambda}{j} \hat{J}_z^2 + \alpha_1 \hat{J}_x  \hat{\sigma}_x \sum_n \delta \left [ t- nT\right ] \nonumber \\
&&+ \alpha_2  \hat{J}_y \hat{\sigma}_y \sum_n \delta \left [ t -(n+1/2) T \right ] 
\label{eq:topham}
\end{eqnarray}
where $\hat{J}_a$, $a = x,y,z$ are the top operators obeying the usual commutation relation $[\hat{J}_i, \hat{J}_j] = i \epsilon_{ijk} \hat{J}_k$ and the Pauli matrices represent the spin-1/2 degrees of freedom which we label as $\uparrow$ and $\downarrow$.  The parameter $\Lambda$ is the nonlinear energy, and $\alpha_1$ and $\alpha_2$ are the first and second kick strengths, respectively.  Each period $T$ there are two kicks where the second kick is delayed by a time $T/2$. 

The dynamics generated after one period is given by the Floquet operator

\begin{eqnarray}
\hat{U}_T = e^{-i \Lambda \hat{J}_z^2 T/2j} e^{-i\alpha_2  \hat{J}_y \hat{\sigma}_y }e^{-i \Lambda \hat{J}_z^2 T/2j} e^{-i\alpha_1 \hat{J}_x \hat{\sigma}_x }\, .\nonumber \\
\end{eqnarray}
We analyze a simplified version of $\hat{U}_T$ by setting the period to $T = 4 \pi j/\Lambda$.  This is similar to the on-resonance condition found in variations of the quantum kicked rotor \cite{talukdar10,kanem07,ullah11} and results in the unitary operators containing $\hat{J}_z^2$ becoming unity since the eigenvalues of $\hat{J}_z$, $m$, are integers in the range $-j \leq m \leq j$.  The Floquet operator becomes 

\begin{equation}
\hat{U}_T =  e^{-i\frac{\kappa_2}{j}\hat{J}_y \hat{\sigma}_y } e^{-i\frac{\kappa_1}{j} \hat{J}_x \hat{\sigma}_x }
\label{eq:top1}
\end{equation}
where $\kappa_1 = \alpha_1  j$ and $\kappa_2 = \alpha_2  j$ are the scaled kick strengths. 

If the interactions between the top and the spin-1/2 degrees of freedom in Eq.\ \eqref{eq:top1} are difficult to generate, then a possible solution is to use interactions of the general form $\hat{H}_\mathrm{int} = \hat{J}_z \hat{S}_z$, where in our case $\hat{S}_z = \hat{\sigma}_z$.   They can arise in squeezing experiments involving interactions between matter \cite{huang21} or between matter and light \cite{bao20}.  Also, similar interactions are found in a Bose-Fermi mixture in an optical lattice \cite{will11}.  Using $\hat{H}_\mathrm{int}$ as a starting point, we can construct $\hat{U}_T$ from the rotation operators $\hat{R}_x(a) = e^{-i a \hat{J}_x}$, $\hat{R}_y(a) = e^{-i a \hat{J}_y}$ and $\hat{R}_z(a) = e^{-i a \hat{J}_z}$  and the general Pauli rotation operator

\begin{equation}
\hat{M}(\alpha, \beta) = \begin{pmatrix}
\cos(\alpha/2) & \sin(\alpha/2) e^{- i \beta} \\
 -\sin(\alpha/2) e^{ i \beta} & \cos(\alpha/2)
\end{pmatrix} \, .
\end{equation}
The $x$ and $z$ rotation operators can be thought of as a phase accumulation when only tunneling and only an imbalance between the two modes is switched on, respectively, and can be used to create the $y$ rotations since $\hat{R}_y(a) = \hat{R}_z(\pi/2) \hat{R}_x(a) \hat{R}_z(-\pi/2)$.  The explicit breakdown of the unitary operators in $\hat{U}_T$ is

\begin{eqnarray}
 e^{-i \kappa_1 \hat{J}_x\hat{\sigma}_x/j} &=& \hat{M}(-\pi/2,0) \hat{R}_y(\pi/2) e^{-i \kappa_1 \hat{J}_z\hat{\sigma}_z/j} \nonumber \\
&& \hspace{20pt} \times \hat{R}_y (-\pi/2) \hat{M}(\pi/2,0) \nonumber \\
 e^{-i \kappa_2 \hat{J}_y\hat{\sigma}_y/j} &=& \hat{M}(-\pi/2,\pi/2) \hat{R}_x(-\pi/2) e^{-i \kappa_2 \hat{J}_z\hat{\sigma}_z/j} \nonumber \\
&& \hspace{20pt} \times \hat{R}_x(\pi/2) \hat{M}(\pi/2,\pi/2) \, .
\end{eqnarray}

A recent proposal of a quantum walk on the Bloch sphere \cite{duan22} used the Floquet operator 

\begin{equation}
\hat{U}_W = e^{-i2\kappa \hat{J}_z \hat{\sigma}_z} \hat{M}(\alpha, \beta) 
\end{equation}
to evolve the system.  In this context, the Pauli operators represent a quantum coin which is tossed at each step via $\hat{M}(\alpha, \beta)$, then a rotation about the $J_z$ axis is performed whose direction depends on the state of the coin.  Therefore, $\hat{U}_T$ can be thought of as the Floquet operator for a quantum walk on the Bloch sphere involving four coin tosses and two rotations, one about the $J_y$ axis and one about the $J_x$ axis, at each step. 

Going forward we will discuss topological regions that emerge in the space of $\hat{J}_z$ eigenstates, $\{\vert m \rangle \}$.  To distinguish between the different topological regions we use terms which are commonly found in condensed matter physics such as 'bound' and 'bulk' to describe the states located near and away from the boundaries separating the regions, respectively.  We also use the term 'edge' to refer to the minimum and maximum angular momentum states of the top, $\vert m = \pm j\rangle$.  Therefore, it is useful to consider $m$ as the spatial coordinate of a fictitious 1D lattice.   An important property of this lattice is that it does not possess discrete translational symmetry which can be seen from the raising and lowering operators of the top

\begin{equation}
\hat{J}_\pm \vert m \rangle  = \sqrt{(j\mp m)(j\pm m+1)} \vert m \pm 1 \rangle \, .
\label{eq:pm}
\end{equation}
The inhomogeneity comes from the Hilbert space of the top which is a Bloch sphere of radius $R = \sqrt{j(j+1)}$.  In order to see this explicitly one can imagine a system consisting of a single particle in a 1D lattice, with nearest neighbor hopping, that has been stretched on a sphere from the north pole to the south pole. If the tunneling energy is inversely proportional to the distance between sites the Hamiltonian is

\begin{equation}
\hat{H}_\mathrm{sph} = -\alpha j \sum_m \frac{1}{d_{m,m+1}} \left (\hat{a}_{m+1}^\dagger \hat{a}_m + \mathrm{h.c.} \right ).
\end{equation}
Due to the curved surface of the sphere, the distance between adjacent sites is simply the arclength connecting them $d_{m,m+1} = R \vert \Delta \theta_{m,m+1}\vert$ where $ \Delta \theta_{m,m+1} = \theta_m - \theta_{m+1}$ is the difference between the polar angle coordinates of the two sites.  To make a connection to the $\hat{J}_z$ states of the top, we require that the site label $m$ also be the polar axis coordinate and that it takes unit increments in the range $-j \leq m \leq j$.  This gives the relation $m = j \cos\theta$ or $\theta_m = \arccos (m/j)$. Finally, we assume a large system size ($j \gg 1$), so that $R \approx j$ and take the leading order term in a $1/j$ expansion of the distance $d_{m,m+1} \approx \left [1-(m/j)^2\right ]^{-1/2}$ which gives

\begin{equation}
\hat{H}_\mathrm{sph} \approx -\alpha  \sum_m \sqrt{j^2 - m^2} \left (\hat{a}_{m+1}^\dagger \hat{a}_m + \mathrm{h.c.} \right ) \, .
\label{eq:Hsp}
\end{equation}

Although Eq.\ \eqref{eq:Hsp} is an approximation, the square root factor is the mean field version of the ones in Eq.\ \eqref{eq:pm}, so their difference relative to $j$ vanishes as $j \to \infty$.  Here, we see $\hat{J}_+$ has a similar to the single particle tunneling terms  $\sum_m \sqrt{j^2 - m^2} \hat{a}_{m+1}^\dagger \hat{a}_m$.  This quick analysis is not meant to discuss how $\hat{H}_\mathrm{sph}$ can be implemented, but rather highlight two main points: (1) the eigenvalues of $\hat{J}_z$ are similar to the site label of a 1D lattice stretched over the semicircle connecting the two poles of a sphere and (2) the square root factors come from the curvature of the sphere and therefore have a geometric origin.  We stress the second point because, as we will show, it is the square root factors that are responsible for the breakdown of the state space of the spin-1/2 QDKT into regions of different topology and therefore the boundaries between these regions are of geometric origin.

\section{Results}

\subsection{Quasienergy spectrum}

When dealing with time periodic systems it is convenient to use Floquet theory which allows one to write the dynamics over one period in terms of a time independent effective Hamiltonian 

\begin{equation}
\hat{U}_T =  e^{-i \hat{H}_\mathrm{eff} T} .
\label{eq:top2}
\end{equation}
The set of eigenvalues of the Floquet operator are $\{\lambda_i \}$ and they can be used to calculate the eigenvalues of the effective Hamiltonian $\{\varepsilon_i\} = \{ \frac{\mathrm{i}}{T} \mathrm{log} \lambda_i \}$, however, they are only unique within a range of $2\pi$, so they are referred to as quasienergies.  Going forward we set $T = 1$ without loss of generality.  Before we discuss the quasienergy spectrum we will briefly go over some subtleties in quantifying the topology of Floquet systems.

\begin{figure}[t]
\centering
\includegraphics[scale=0.65]{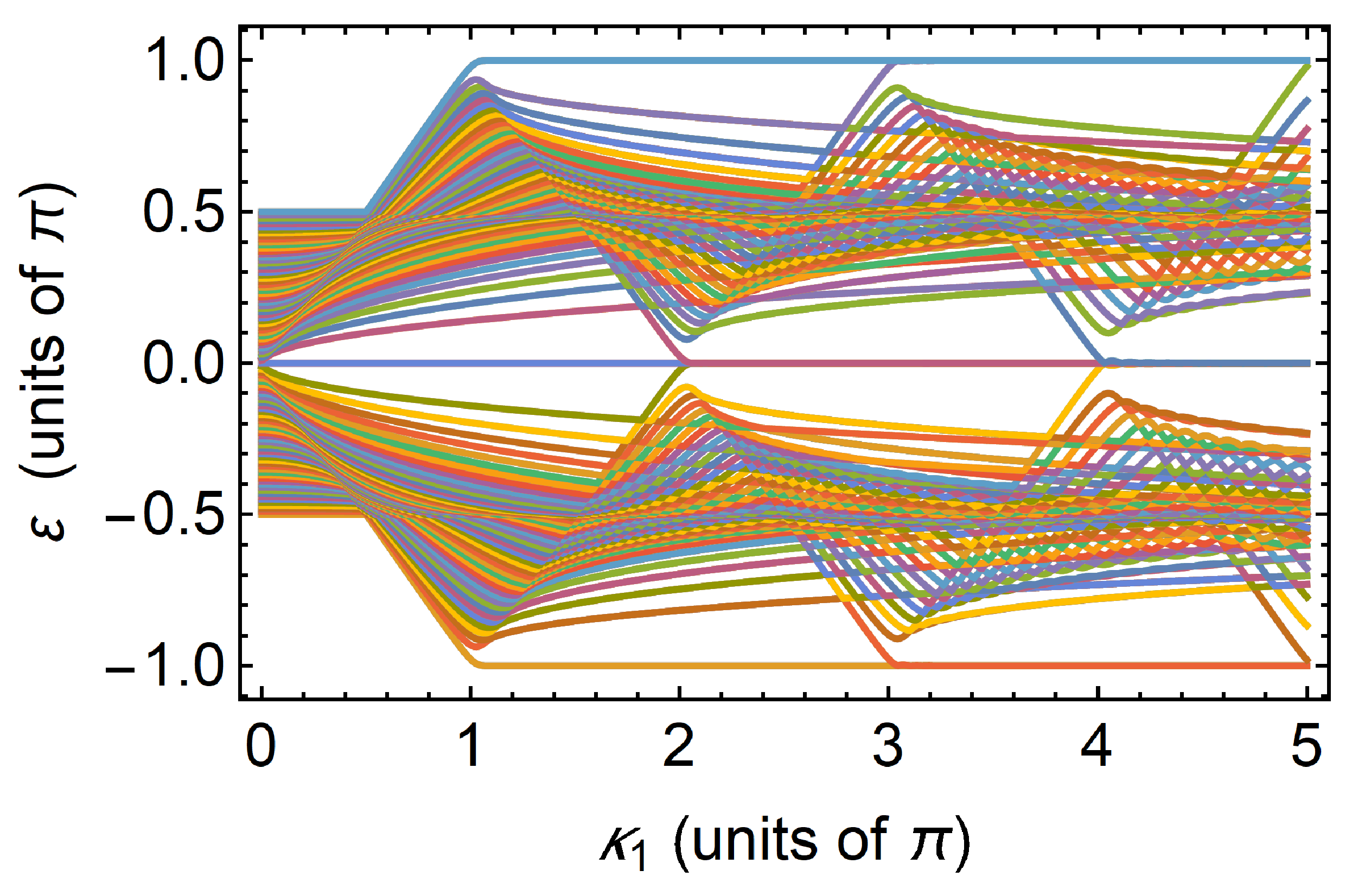}
\caption{Quasienergies as a function of kick strength.  Eigenvalues of $\hat{U}_T$ in Eq.\ \eqref{eq:top1} as a function of $\kappa_1$ for a fixed value of $\kappa_2 = 0.5 \pi$ and $j = 50$.  When $\kappa_1/\pi$ is even or odd a new pair of $\varepsilon = 0$ or $\varepsilon = \pi$ states are formed, respectively.}
\label{fig:QEn}
\end{figure}

In static systems, topological phases are characterized by integers such as the winding number.  Through the bulk-boundary correspondence, they count the number of protected bound states at the boundary of the bulk.  In periodically driven systems it has been shown that, in addition to the usual $\varepsilon  = 0$  bound states, there are also $\varepsilon  = \pi$ bound states which come from the fact that the quasienergies are calculated from a unitary operator and not a Hamiltonian.  These states are protected \cite{roy17} and cannot be deformed into each other without an energy gap closing, or breaking of some symmetry, so two winding numbers are required to characterize each phase.  Care must be taken in calculating these numbers for Floquet systems, however, and it has been shown that the winding numbers of two chiral symmetrized timeframes which are $w_1$ and $w_2$, can be used to calculate the winding numbers that count the number of $\varepsilon =0 $ and $\varepsilon = \pi$ bound states from the relation \cite{asboth12}

\begin{equation}
w_0 = \frac{w_1 + w_2}{2} \hspace{30pt} w_\pi = \frac{w_1 - w_2}{2} \, .
\end{equation}
In the spin-1/2 QDKT the Floquet operators in the chiral symmetrized timeframes take the form

\begin{eqnarray}
\hat{U}_{T,1} &=&  e^{-i\frac{\kappa_1}{2j}\hat{J}_x \hat{\sigma}_x } e^{-i\frac{\kappa_2}{j} \hat{J}_y \hat{\sigma}_y } e^{-i\frac{\kappa_1}{2j}\hat{J}_x \hat{\sigma}_x }  \label{eq:UT1} \\
\hat{U}_{T,2} &=&  e^{-i\frac{\kappa_2}{2j}\hat{J}_y \hat{\sigma}_y } e^{-i\frac{\kappa_1}{j} \hat{J}_x \hat{\sigma}_x } e^{-i\frac{\kappa_2}{2j}\hat{J}_y \hat{\sigma}_y } \label{eq:UT2} \, .
\end{eqnarray}
The chiral symmetry that Eqns.\ \eqref{eq:UT1} and \eqref{eq:UT2} possess is defined in terms of the relations $\hat{\Gamma} \hat{U}_{T,1} \hat{\Gamma} =  \hat{U}_{T,1}^\dagger$ and $\hat{\Gamma} \hat{U}_{T,2} \hat{\Gamma} =  \hat{U}_{T,2}^\dagger$ for the operator $\hat{\Gamma} = \hat{\sigma}_z$.  This means that for any state with quasienergy $\varepsilon$ there is a partner state with quasienergy $-\varepsilon$ through the relation $\hat{\sigma}_z \vert \varepsilon \rangle  = \vert -\varepsilon\rangle$.  The Floquet operators in Eqns. \eqref{eq:top1}, \eqref{eq:UT1} and \eqref{eq:UT2} are separated by unitary transformations, so they have the same spectrum.  Figure \ref{fig:QEn} shows the spectrum for a fixed value of $\kappa_2 = 0.5 \pi$ and variable $\kappa_1$ for $j = 50$.  Increasing $\kappa_1$ creates new $\varepsilon = 0$ and $\varepsilon = \pi$ states at even and odd integer multiples of $\pi$, respectively.  These states are the bound states mentioned earlier and are protected in the sense that one can perturb the Floquet operators with a general term in the exponentials that respects the chiral symmetry (terms proportional to $\hat{\sigma}_x$ or $\hat{\sigma}_y$) without changing the number of bound states.

We note that the bound states do not have quasienergies $\varepsilon = 0$ and $\varepsilon = \pi$ exactly due to finite size effects.  Evidence of this can be seen around $\kappa_1 = 4 \pi$ where there is a slight wiggle in the energy around $\varepsilon = 0$.  Nevertheless, for the range of values and system size shown, the bound states are quite stable.  Additionally, we note that the form of the spectrum depends on the value of $\kappa_2$ and for general values the number of bound states does not always increase monotonically as $\kappa_1$ increases.  This is highlighted in the phase diagram of the spin-1/2 QDKR \cite{zhou18} which quickly becomes complicated as $\kappa_1$ and $\kappa_2$ increase.  With the intent of keeping our analysis simple, we set $\kappa_2 = 0.5 \pi$ for the remainder of the paper.

\subsection{Locations of bound states}

To gain a better insight into the bound states as well as the topology of the system it would be useful to calculate $\hat{H}_\mathrm{eff}$ for the chiral symmetrized Floquet operators in Eqns.\ \eqref{eq:UT1} and \eqref{eq:UT2}, however, their forms are not obvious.  Some progress can be made by performing a mean field approximation of the top operators by transforming them into their coherent state expectation values

\begin{equation}
\langle \hat{\boldsymbol{J}} \rangle = \langle  ( \hat{J}_x, \hat{J}_y, \hat{J}_z ) \rangle = j (\sin\theta \cos \phi, \sin\theta \sin\phi, \cos\theta) \, . 
\end{equation}
The angles $\theta$ and $\phi$ are the polar and azimuthal angles, respectively, of the Bloch sphere of the top with radius $R = \sqrt{j(j+1)}$.  Instead of the polar angle, we find it useful to use the eigenvalue label of $\hat{J}_z$, $m$, as a variable through the previously mentioned relation $m = j\cos\theta$.  Defining the new parameters 

\begin{eqnarray}
K_1 &=& \kappa_1 \sqrt{1-(m/j)^2} \cos\phi \nonumber \\
K_2 &=& \kappa_2 \sqrt{1-(m/j)^2} \sin\phi \, ,
\label{eq:Kpar}
\end{eqnarray}
the mean field versions of Eqns.\ \eqref{eq:UT1} and \eqref{eq:UT2} become

\begin{eqnarray}
\hat{U}_{T,1}^\mathrm{MF} &=&  e^{-i\frac{K_1}{2} \hat{\sigma}_x } e^{-iK_2 \hat{\sigma}_y } e^{-i\frac{K_1}{2}\hat{\sigma}_x }  \label{eq:UT1MF} \\
\hat{U}_{T,2}^\mathrm{MF} &=&  e^{-i\frac{K_2}{2} \hat{\sigma}_y } e^{-iK_1 \hat{\sigma}_x } e^{-i\frac{K_2}{2} \hat{\sigma}_y } \label{eq:UT2MF} \, .
\end{eqnarray}
In this form, the effective Hamiltonians can be calculated exactly giving $\hat{H}_{\mathrm{eff},i}^\mathrm{MF} = \varepsilon \boldsymbol{n_i} \cdot \hat{\boldsymbol{\sigma}}$ (Appendix A), where $i = 1, 2$ for the two symmetrized timeframes.  Both the quasienergy $\varepsilon$ and the vector $\boldsymbol{n_i} = (n_{ix},n_{iy})$, which has unit length, depend on the state of the top.  This means $\boldsymbol{n_i}$  maps points on the Bloch sphere of the top to points on the equator of the spin-1/2 Bloch sphere.  


The quasienergy is the same for both timeframes in Eqns.\ \eqref{eq:UT1MF} and \eqref{eq:UT2MF} and is

\begin{equation}
\varepsilon = \arccos \left[ \cos(K_1)\cos(K_2)\right ] \, .
\label{eq:EN}
\end{equation}
Topological transitions in Floquet systems are marked by bound states which close the quasienergy gaps at $\varepsilon = 0$ and $\varepsilon = \pi$.   Looking at Eq.\ \eqref{eq:EN}, we see that these energies occur when $\cos(K_1)\cos(K_2) = \pm 1$ which in turn occurs when $\kappa_1 \sqrt{1-(m/j)^2} \cos\phi = \mu \pi$ and $\kappa_2 \sqrt{1-(m/j)^2} \sin\phi = \nu \pi$, where $\mu$ and $\nu$ are integers.  Combining these two results gives

\begin{equation}
\frac{\pi^2}{1-(m/j)^2} \left [\frac{\mu^2}{\kappa_1^2} + \frac{\nu^2}{\kappa_2^2} \right ] = 1 \, .
\label{eq:PB}
\end{equation}
which we use to find the mean field values of $m$ that correspond to the quasienergies $\varepsilon = 0$ and $\varepsilon = \pi$

\begin{equation}
m_{\mu,\nu} = \pm j \sqrt{1-\pi^2 \left (\frac{\mu^2}{\kappa_1^2} + \frac{\nu^2}{\kappa_2^2} \right )} \, .
\label{eq:PB1}
\end{equation}
Equation \eqref{eq:PB1} is the mean field prediction of the locations of localized bound states and consequently gives the locations of the boundaries separating different topological regions.  The source of these boundaries comes from the mean field square root factor, $\sqrt{1-(m/j)^2}$, which is due to the curvature of the Bloch sphere.  The geometric origins of these boundaries is reminiscent of geometry induced domain walls which appear in a 2D lattice of dipoles on the surface of a torus \cite{siemens23}. 

\begin{figure}[t]
\centering
\includegraphics[scale=0.65]{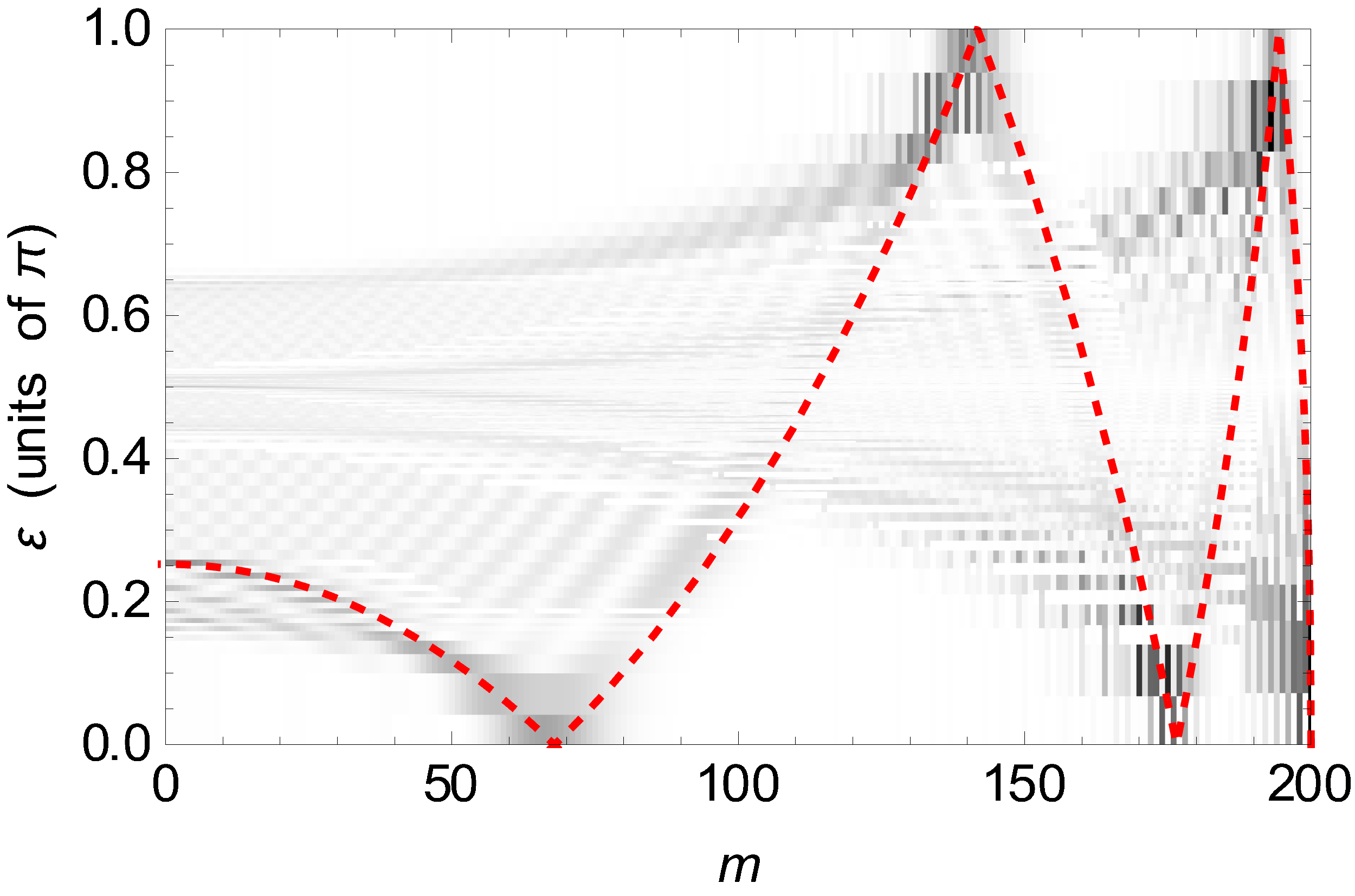}
\caption{Density plot of the probability for a given eigenstate of the top to have a given quasienergy.  Darker shades of black show higher values of the probability $\vert \langle m \vert \varepsilon_i \rangle\vert^2$  where the $\hat{J}_z$ eigenstate label, $m$, and the quasienergy, $\varepsilon_i$, vary along the $x$ and $y$ axes, respectively.  The red dashed curve is the mean field quasienergy from Eq.\ \eqref{eq:EN} for $\phi =0$.  The points of interest are the kinks in the mean field result at $\varepsilon = 0, \pi$ where protected edge states are shown in the density plot as localized states around these points.  The parameter values are $\kappa_1 = 4.25 \pi$, $\kappa_2 = 0.5 \pi$ and $j = 200$.}
\label{fig:denstop}
\end{figure}

To test the mean field predictions we calculate the probability for a given $\hat{J}_z$ state to be in an eigenstate of the Floquet operator in Eq.\ \eqref{eq:top1}, $P(m, \varepsilon_i) = \vert \langle m \vert \varepsilon_i \rangle \vert^2$.  Figure \ref{fig:denstop} shows a density plot of the probability with the $m$ state label and the quasienergy on the $x$ and $y$ axes, respectively, along with the $m$ dependence of the mean field quasienergy for the azimuthal angle $\phi = 0$ (dashed, red).  The kinks in the mean field result at  $\varepsilon = 0$ and $\varepsilon = \pi$ are the locations of the localized bound states predicted from Eq.\ \eqref{eq:PB1} and they agree quite well with the quantum probability which shows localized states around these points.  The creation (destruction) of bound states takes place at the equator of the Bloch sphere at $m = 0$.  This means that as $\kappa_1$ decreases in Fig.\ \ref{fig:denstop}, the $\varepsilon = 0$ bound states at $m \approx 67$ will move toward $m = 0$ until they reach that point when $\kappa_1 = 4 \pi$.  Further decreasing $\kappa_1$ results in the bound states' destruction as they are absorbed into the bulk.



%

%

\subsection{Local winding numbers}

The boundaries partition the Bloch sphere into regions of different topology which are quantified in terms of winding numbers.  Under the mean field approximation the winding number has a simple geometric interpretation as the number of times the vector  $\boldsymbol{n_i}$ winds around the origin as $\phi$ goes from $-\pi$ to $\pi$.  However, due to the square root factors, the winding number is $m$ dependent, so we calculate local versions of them.   The mean field local winding number is calculated with the equation \cite{cardano17}

\begin{equation}
w_{\mathrm{MF},i}(m) = \int_{-\pi}^\pi \frac{d\phi}{2\pi} \boldsymbol{n_i} \times \partial_\phi \boldsymbol{n_i} \, .
\label{eq:MFWN}
\end{equation}
The quantum winding number calculation must also be performed locally due to $\phi$ not being good quantum number.  The calculation relies on the flat-band transformation of the  effective Floquet Hamiltonian, $\hat{Q} = \hat{P}_+ - \hat{P}_-$, where $\hat{P}_+$ is the projector onto eigenstates with $\varepsilon >0$ and $\hat{P}_-$ is the projector onto eigenstates with $\varepsilon <0$.  The chiral symmetry of $\hat{H}_\mathrm{eff}$ allows us to write the flat-band projector as $\hat{Q}=\hat{Q}_{\uparrow\downarrow} + \hat{Q}_{\downarrow\uparrow}$ where $\hat{Q}_{\uparrow\downarrow}  = \hat{\Gamma}_\uparrow \hat{Q} \hat{\Gamma}_\downarrow$ and $\hat{\Gamma}_\uparrow$, $\hat{\Gamma}_\downarrow$ are the projectors onto the spin-1/2 $\uparrow$ and $\downarrow$ states, respectively.  The spin-1/2 projectors form the chiral symmetry operator $\hat{\Gamma} = \hat{\Gamma}_\uparrow - \hat{\Gamma}_\downarrow$ and the symmetrized winding number operator is then \cite{song14,shem14,meier18}

\begin{equation}
\hat{w}_i =  \left (\hat{Q}_{\downarrow\uparrow,i} [ \hat{J}_z, \hat{Q}_{\uparrow\downarrow,i}]+\hat{Q}_{\uparrow\downarrow,i} [ \hat{J}_z, \hat{Q}_{\downarrow\uparrow,i}] \right )/2
\label{eq:QWN}
\end{equation}
where the index $i=1, 2$ is for the two chiral timeframes from Eqns.\ \eqref{eq:UT1} and \eqref{eq:UT2}.  Therefore, the winding number localized to a single $\hat{J}_z$ state is $w_i(m) = \mathrm{Tr}_\sigma \langle m \vert \hat{w}_i \vert m \rangle$ where the trace is over the spin-1/2 degrees of freedom.  It is expected that for large system sizes and for states comfortably within a bulk (away from the topological boundaries), the quantum and mean field calculations will agree.

\begin{figure}[t]
\centering
\includegraphics[scale=0.18]{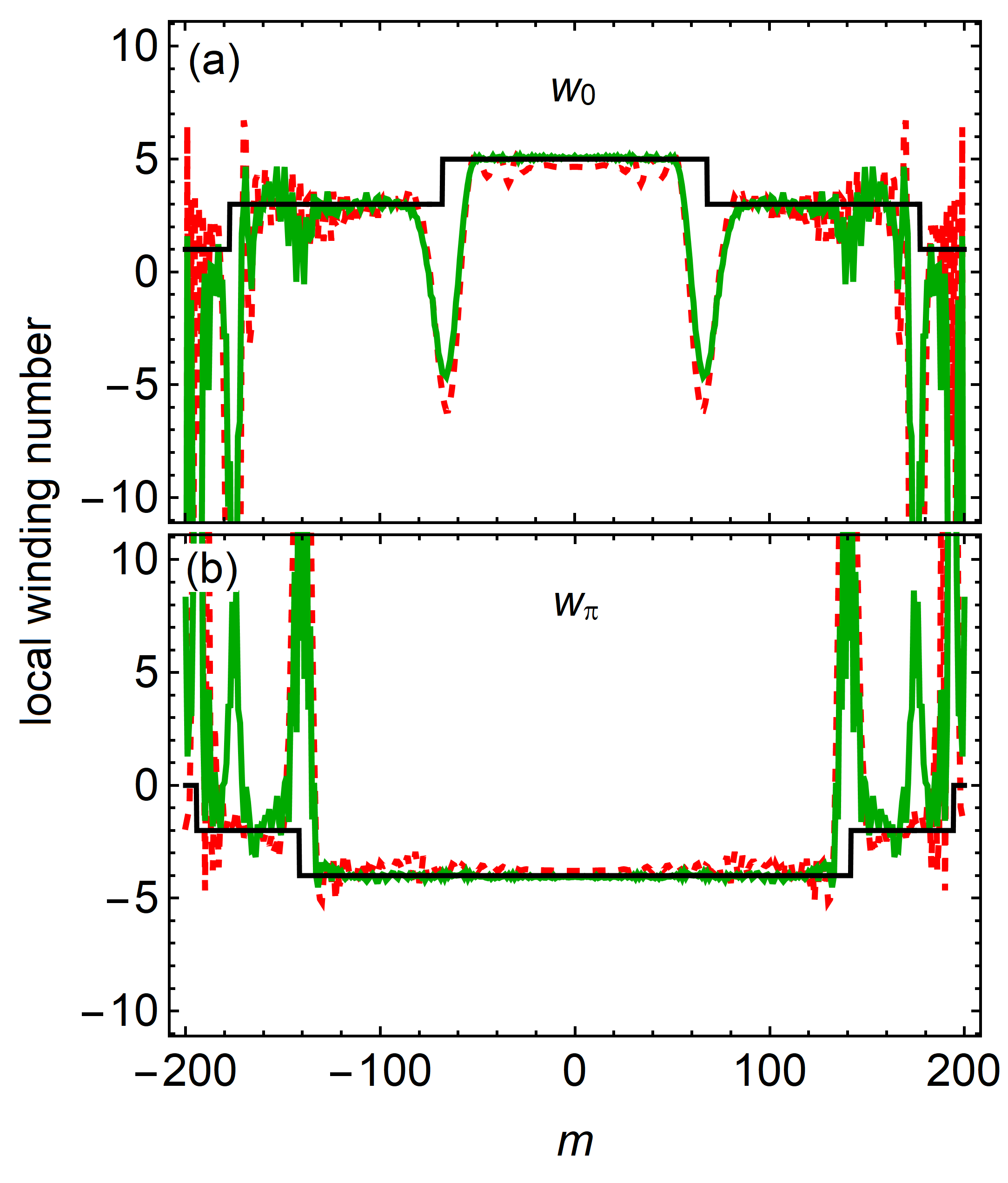}
\caption{Local winding numbers.  (a) Comparison of the mean field (solid black), quantum (solid green) and mean chiral displacement (dashed red) predictions of the $\varepsilon = 0$ local winding number over the space of $\hat{J}_z$ states for $\kappa_1 = 4.25 \pi$, $\kappa_2 = 0.5 \pi$ and $j = 200$.  The locations of the steps in the mean field prediction can be calculated from Eq.\ \eqref{eq:PB1}.  (b) Same as (a) except the calculations are performed for the $\varepsilon = \pi$ local winding number.}
\label{fig:WN}
\end{figure}

In Fig.\ \ref{fig:WN} (a) and (b), we plot the $m$ dependence of the mean field (black) and quantum (green) values of $w_0$ and $w_\pi$, respectively, for $\kappa_1 = 4.25 \pi$ and $\kappa_2 = 0.5 \pi$.  The different topological regions are shown as plateaus while the boundaries are shown as a steps in the mean field case and jumps/dips in the quantum case.  In each image the step locations are calculated from Eq.\ \eqref{eq:PB1} $m_{\mu, \nu} =\pm j\sqrt{1-16\mu^2/289 - 4\nu^2}$ where we find five pairs of integers which satisfy the equation: $0\leq\mu\leq4$ and $\nu = 0$.  The quantum local winding number agrees with the mean field result strongly in the central bulk and the agreement falls off moving closer to the edge at $m = \pm 200$.  This is expected since finite size effects become pronounced at the edges of the system.  Another obvious feature of the quantum local winding number is the sudden jumps/dips at the boundaries between different topological regions.  These are attributed to the fact that the boundaries are home to exponentially localized bound states, so they are close to being eigenstates of $\hat{J}_z$.  

Looking at Fig.\ \ref{fig:QEn} we see that the parameter values of $\kappa_1 = 4.25 \pi$ and $\kappa_2 = 0.5 \pi$ mean there are five $\varepsilon = 0$ states and four $\varepsilon = \pi$ states.  In Fig.\ \ref{fig:WN}, the quantum local winding numbers for the central region are $\vert w_0 \vert \approx 5$ and $\vert w_\pi \vert \approx 4$ which matches the number of bound states and therefore supports the bulk-boundary correspondence.  To determine how the outer regions support the bulk-boundary correspondence we once again turn to the mean field results, specifically the parameters in Eq.\ \eqref{eq:Kpar}.  There we see that moving away from the equator of the Bloch sphere at $m = 0$ has the same effect on $K_1$ and $K_2$ as decreasing the scaled kick strengths $\kappa_1$ and $\kappa_2$.  Figure  \ref{fig:QEn} shows that as $\kappa_1$ decreases, bound states are destroyed which is why we see the mean field local winding number decrease in steps in Fig.\ \ref{fig:WN} as one moves toward the edges at $j = \pm 200$.  We see the same decrease in the quantum local winding numbers, however, the fluctuations increase as the edge is approached due to finite size effects. 


The winding number is not always directly measurable in experiments and it is often easier to extract information from the dynamics of the system.  To this end, the chiral displacement  \cite{cardano17,maffei18}

\begin{eqnarray}
C_i(m,n) &=&\mathrm{Tr}_\sigma \langle m \vert \hat{U}_{T,i}^{-n} \hat{J}_z \hat{\sigma}_z \hat{U}_{T,i}^{n} \vert m \rangle \nonumber \\ 
 &=& \left ( \sum_{a = \uparrow, \downarrow} \sum_{j,k} c_{j, a}^{m*} c_{k,a}^m e^{-i ( \varepsilon_j - \varepsilon_k) n} \langle \varepsilon_k\vert \hat{J}_z  \hat{\sigma}_z \vert \varepsilon_j \rangle \right )_i \nonumber \\
\label{eq:CD}
\end{eqnarray}
has been shown to be related to the quantum winding number.  In the second line of Eq.\ \eqref{eq:CD}, we take the spectral decomposition of the Floquet operators, $c_{j,a}^m = \langle a, m \vert \varepsilon_j \rangle$ and the subscript $i = 1,2$ indicates which Floquet operator we are using from Eqns.\ \eqref{eq:UT1} and \eqref{eq:UT2}.  The relation to the winding number comes from the fact that the local winding number at $m = 0$ can be approximated as \cite{meier18}

\begin{eqnarray}
w_i(0) &=& \mathrm{Tr}_\sigma \langle 0 \vert \hat{w}_i \vert 0 \rangle \nonumber \\
&=& \left ( \sum_{a = \uparrow, \downarrow} \sum_{j,k} c_{j, a}^{0*} c_{k,a}^0  \langle \varepsilon_k\vert \hat{J}_z  \hat{\sigma}_z \vert \varepsilon_j \rangle \right )_i \, . \nonumber \\
\end{eqnarray}
Comparing this equation with Eq.\ \eqref{eq:CD}, we see that a long time average of the chiral displacement at $m=0$ 

\begin{equation}
\overline{C_i(0)} = \lim_{\mathcal{N}\to\infty} \frac{1}{\mathcal{N}} \sum_n^\mathcal{N} C_i(0,n)
\label{eq:TA}
\end{equation}
will be equal to the winding number if the off-diagonal terms of $\langle \varepsilon_j \vert \hat{J}_z\hat{\sigma}_z \vert \varepsilon_k \rangle$ can be neglected since they get 'washed-out' in the averaging process.  This was found to be the case for the spin-1/2 QDKR \cite{zhou18} and in a synthesized 1D topological wire based on the momentum states of a BEC \cite{meier18}.  Although these terms are small for our system, they are non-negligible, so they need to be included in order for the dynamics to display an accurate winding number.  Therefore, short time averages are better in order for the off-diagonal terms to not vanish completely.  The dashed red curves in Fig.\ \ref{fig:WN} (a) and (b) show the chiral displacement averaged over $\mathcal{N} = 20$ steps for all states (not just the $m=0$ state).  We see it maintains a similar shape to the local winding number calculated from Eq.\ \eqref{eq:QWN} (green) including the sudden jumps/dips at the boundaries, however, it does fall short of the mean field and quantum local winding numbers over the majority of the states due to the averaging process.  A comparison between short and long time averages of the chiral displacement can be found in Appendix \ref{app:LTA}.

\subsection{Using a probe state to locate topological boundaries}

\begin{figure*}[t]
\centering
\includegraphics[scale=0.47]{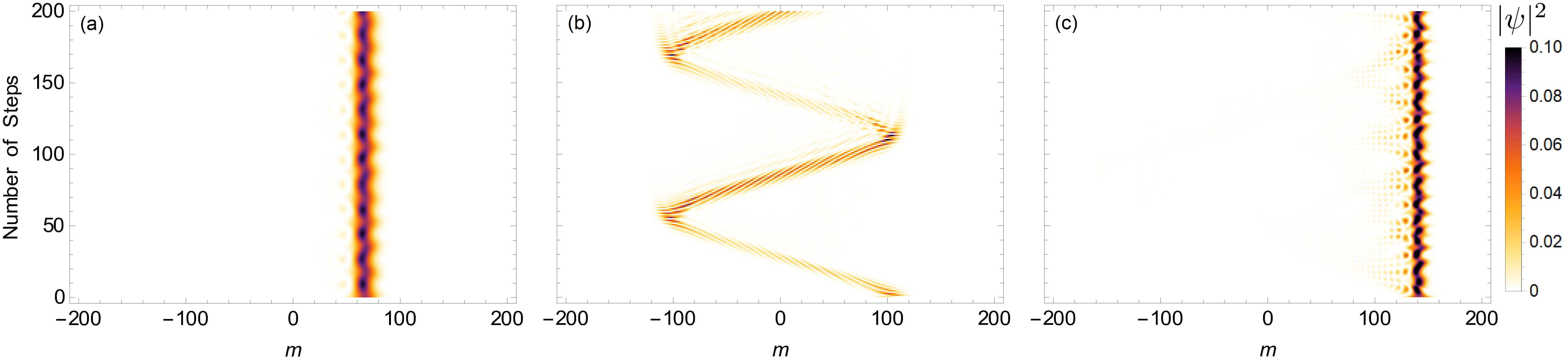}
\caption{Evolution of probability distributions for different initial states. (a) Evolution generated by $\hat{U}_T$ in Eq.\ \eqref{eq:top1} of the probability distribution of an initial Gaussian state centered on the boundary between two different topological regions.  The boundary is located near $\lfloor m_{4,0} \rfloor = 67$ and corresponds to the right closest step to the center in Fig.\ \ref{fig:WN} (a), so it is the boundary between the $(w_0,w_\pi) = (5,-4)$ and the $(w_0,w_\pi) = (3,-4)$ topological regions.  (b) Same initial Gaussian state as in (a) except it is centered at $\lfloor m_{4,0} + m_{3,0} \rfloor/2  = 104$ which is halfway between the boundary separating the  $(w_0,w_\pi) = (5,-4)$ and the $ (3,-4)$ regions and the boundary separating the $(w_0,w_\pi) = (3,-4)$ and the $ (3,-2)$ regions.  (c) Same initial Gaussian state as in (a) except it is centered at $\lfloor m_{3,0} \rfloor  = 141$ which is the boundary  separating the  $(w_0,w_\pi) = (3,-4)$ and the $ (3,-2)$ regions   which is the right closest step to the center in Fig.\ \ref{fig:WN} (b). The parameters are $\kappa_1 = 4.25 \pi$, $\kappa_2 = 0.5 \pi$ and $j = 200$.}
\label{fig:IPR}
\end{figure*}

\begin{figure}[t]
\centering
\includegraphics[width=\columnwidth]{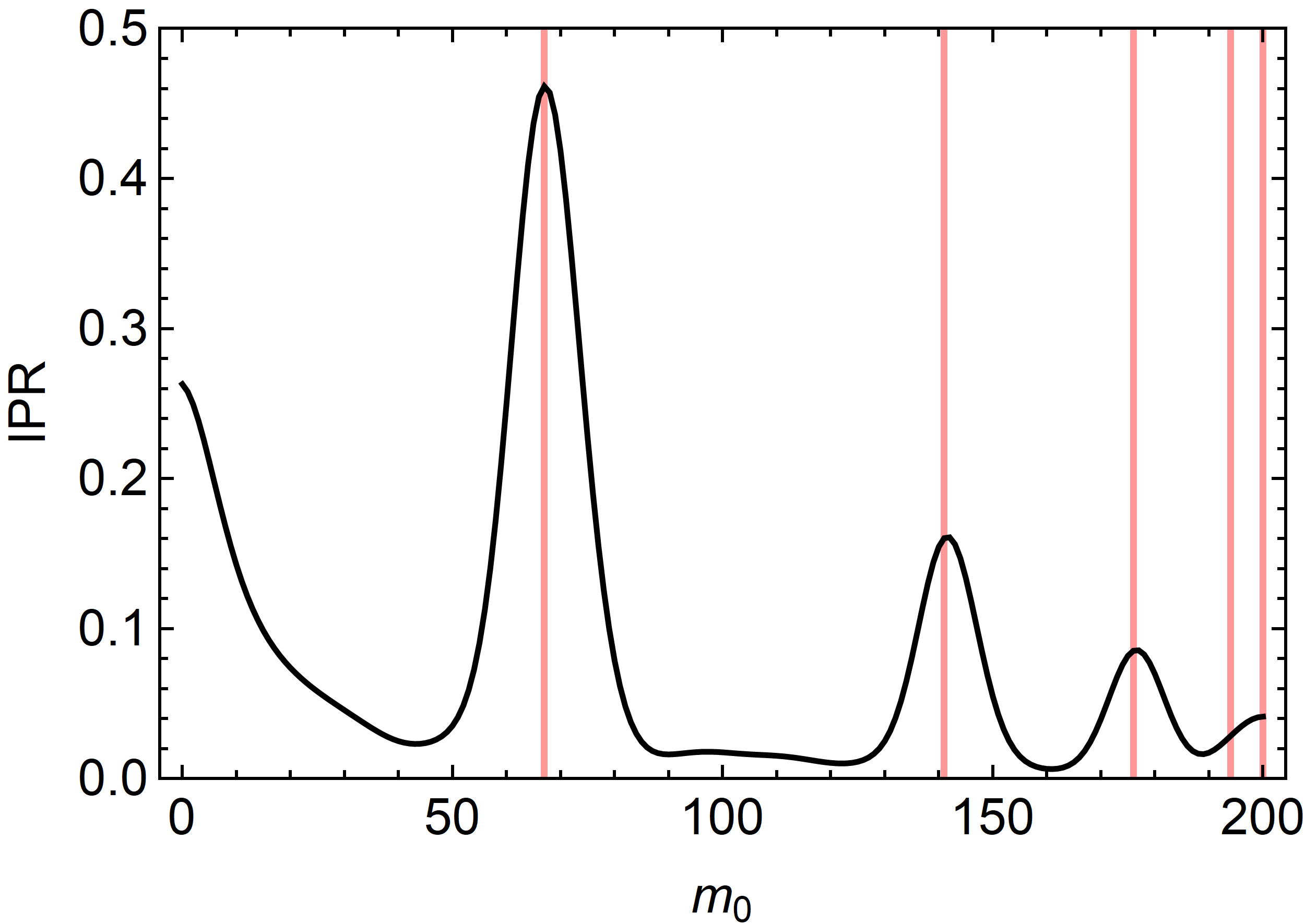}
\caption{Inverse participation ratio of  Gaussian states in the Floquet basis.  The IPR given in Eq.\ \eqref{eq:IPR} of the state   $\vert  \psi_0 \rangle = \sqrt{\frac{1}{\Delta m \sqrt{\pi}}} \sum_m e^{-\frac{(m-m_0)^2}{2\Delta m^2}} \vert m, \uparrow \rangle$ in the basis states of  Eq.\ \eqref{eq:top1} is plotted as a function of $m_0$.  The red vertical lines are the predicted locations of the topological boundaries from Eq.\ \eqref{eq:PB1} which line up with peaks in the IPR where the Gaussian strongly overlaps the bound states located at the boundaries.  The peaks match the locations of the steps in the mean field winding number and the jumps/dips in the quantum winding number in Fig.\ \ref{fig:WN}. The parameter values are $\Delta m = 10$, $\kappa_1 = 4.25 \pi$, $\kappa_2 = 0.5 \pi$ and $j = 200$.  }
\label{fig:IPR2}
\end{figure}

The boundaries separating different topological regions can also be located using another dynamical method which takes advantage of the fact that the boundaries are home to the $\varepsilon = 0$, $\pi$ bound states.  If an initial state has a strong overlap with the bound states at a boundary, then it will remain localized there for a long period of time.  In contrast, an initial state comfortably away from a boundary will explore more of the Hilbert space as it evolves because of the larger overlap with the bulk states.  We choose the initial probe state to be Gaussian in shape $\vert  \psi_0 \rangle = \sqrt{\frac{1}{\Delta m \sqrt{\pi}}} \sum_m e^{-\frac{(m-m_0)^2}{2\Delta m^2}} \vert m, \uparrow \rangle$ with $\Delta m = 10$.  One of the benefits of using this method is that the details of the initial state are not important as long as it can pick out one boundary over another.   In Fig.\ \ref{fig:IPR} (a), we show the evolution of the probability distribution in the $\uparrow$ subspace of the initial Gaussian state centered on the boundary located at $m_0 = \lfloor m_{4,0} \rfloor = 67$.  This boundary is shown in Fig.\ \ref{fig:WN} (a) as the first step away from the $m = 0$ state.  The strong overlap between the initial state and the bound states keeps the distribution localized over the period shown. In Fig.\ \ref{fig:IPR} (b), the initial state is centered at $m_0 = 104$ which is halfway between two boundaries.  Clearly the probability distribution does not remain localized and has a checkered pattern which is due to parts of the wave function occupying the $\downarrow$ subspace which is not shown.  The reason why the checkered pattern appears in (b) and not (a) is because the initial state has a larger overlap with $\varepsilon \neq 0, \pi$ states in (b) and these states have equal support on both subspaces due to chiral symmetry whereas the $\varepsilon = 0, \pi$ bound states have support on a single subspace.  In Fig.\ \ref{fig:IPR} (c), the initial Gaussian is centered on another boundary at $m_0 = \lfloor m_{3,0} \rfloor = 141$ which is the first step from the center in Fig.\ \ref{fig:WN} (b).  Again, the probability distribution is fairly localized with small parts of it propagating away due the Gaussian having a weaker overlap with the bound states compared to the ones in (a).

In order to get a sense of the overall quality of the specific initial state we chose, we take a look at the inverse participation ratio (IPR) which is a good measure of the localization of a state in a given basis      

\begin{equation}
\mathrm{IPR} = \sum_i \vert  \langle \varepsilon_i \vert \psi_0 \rangle \vert^4 \, . 
\label{eq:IPR}
\end{equation}
Here, $\vert \varepsilon_i \rangle$ are the eigenstates of $\hat{U}_{T}$ in Eq.\ \eqref{eq:top1} and $\vert \psi_0\rangle$ is the initial Gaussian state.  The two extreme values are $\mathrm{IPR} = 1$ and $\mathrm{IPR} = \frac{1}{2 (2j+1)}$ when $\vert \psi_0 \rangle$ is completely localized and completely spread in the basis, respectively.  In Fig.\ \ref{fig:IPR2} the IPR is plotted as a function of the center of the Gaussian $m_0$ where the red vertical lines are the locations of the boundaries predicted from Eq.\ \eqref{eq:PB1}.  The peaks correspond to the initial state being localized at the boundaries where it strongly overlaps with the bound states.  The relative height of the peaks, including the peak at $m = 0$, are state dependent, so Gaussians with different widths will produce different qualitative results, however, the peak locations will remain the same.  Nevertheless, the chosen Gaussian does an excellent job at picking out the boundary locations with the only discrepancy being near the edge of the system at $m = 200$.   There the distance between boundaries is similar to $\Delta m = 10$, so the initial state lacks the resolution required to locate those boundaries completely.

\section{Summary and Discussions}

We have shown that, like the spin-1/2 QDKR, the spin-1/2 QDKT has topological regions with large winding  numbers.  However, the two models differ in the number of topological regions for a given pair of kick strengths.  Whereas the spin-1/2 QDKR has a single region, different regions proliferate in the Bloch sphere of the spin-1/2 QDKT as the kick strengths increase.  We quantify the topology of each region by comparing local versions of the mean field, quantum and mean chiral displacement  winding numbers.  We find that away from the edge of the system they agree with the number of bound states at the boundaries separating each region which supports the bulk-boundary correspondence.  Finally, we used a simple dynamical method to locate the boundaries by preparing a Gaussian initial state and evolving it.  When the initial state is centered on a boundary the state remained localized due to the large overlap with the bound states exponentially localized there. 

The spin-1/2 QDKT is a rich topological system with many possible avenues to explore.  As previously mentioned, it is related to quantum walks and can be used to investigate the effects of multiple topological boundaries in the walk space.  A notable departure from former quantum walk studies is the source of the boundaries.  Usually they are implemented via insertion of an inhomogeneity at a specific site, whereas here, they come from the geometry of the Bloch sphere in which the walk is taking place and can therefore be considered as being built-in by nature.  The positions of the boundaries are controlled by the kick strengths which leads to some interesting possibilities for using the topology of the spin-1/2 QDKT as a tool to control the state of the system.  One example of this is in the creation of cat states.  One can imagine initially setting each kick strength close to some multiple of $\pi$, so that a pair of boundaries just forms at the equator of the Bloch sphere of the top ($m = 0$), then preparing the initial product state $\vert \psi_0 \rangle = \frac{1}{\sqrt{2}} (\vert \uparrow \rangle + \vert \downarrow \rangle ) \vert \theta_0 = \pi/2 \rangle$, where $\vert \theta_0 = \pi/2 \rangle$ is a coherent state centered on the equator.  Provided the coherent state has a strong overlap with the bound states at the boundary it will remain clamped there as we showed in Fig.\ \ref{fig:IPR}. If one of the the kick strengths is made to increase slowly in time, then the coherent state is effectively pulled apart with parts of it moving toward opposite poles of the Bloch sphere as the boundaries move away from the equator resulting in the final state $\vert \psi_f \rangle = \frac{1}{\sqrt{2}} (\vert \uparrow, \theta_0 + \theta \rangle + \vert \downarrow, \theta_0-\theta \rangle)$. The final state is often referred to as a Bell-cat state and is used to test entanglement generation and efficient information extraction beyond the original two-qubit Bell state \cite{vlastakis15}.  The process is possible due to the chiral symmetry of the system which forces the bound states at the two newly formed boundaries to have support in opposite spin-1/2 subspaces.

Another topic to explore is the connection between topology and chaos.  Some initial numerical results of the level spacings of the quasienergies indicate that the spin-1/2 QDKT displays chaotic behavior as the kick strengths increase.  It has been shown that nonlinear effects can lead to chaotic behavior in the bulk of a system while the boundaries have topological order \cite{sone22}.  However, the spin-1/2 QDKT is unique in that the number of boundaries also increases as the kick strengths increase.  This raises the question as to how a bulk can be chaotic when boundaries proliferate in the system.  

\appendix

\section{Derivation of winding vector \label{app:vec}}

We can find $\boldsymbol{n_i}$ by expanding the unitary operators in Eqns.\ \eqref{eq:UT1MF} and \eqref{eq:UT2MF} in terms of trigonometric functions using the identity

\begin{equation}
e^{-i a \boldsymbol{b} \cdot \hat{\boldsymbol{\sigma}}} = \cos (a) - i \boldsymbol{b} \cdot \hat{\boldsymbol{\sigma}} \sin (a) 
\end{equation} 
which gives

\begin{eqnarray}
\hat{U}_{T,1}^\mathrm{MF} &=& \cos K_1\cos K_2 - i \left ( \sin K_1 \cos K_2\hat{\sigma}_x  + \sin K_2 \hat{\sigma}_y\right ) \nonumber \\ \\
\hat{U}_{T,2}^\mathrm{MF} &=& \cos K_1\cos K_2 - i \left (\sin K_1 \hat{\sigma}_x  + \sin K_2 \cos K_1 \hat{\sigma}_y\right ) \, . \nonumber \\
\end{eqnarray}
For $\hat{U}_{T,1}^\mathrm{MF}$ we define

\begin{eqnarray}
\cos \varepsilon &=& \cos K_1 \cos K_2 \nonumber \\
\sin \varepsilon &=& \sqrt{\sin^2 K_1 \cos^2 K_2 + \sin^2 K_2} \nonumber
\end{eqnarray}
and similarly for $\hat{U}_{T,2}^\mathrm{MF}$ we define

\begin{eqnarray}
\cos \varepsilon &=& \cos K_1 \cos K_2 \nonumber \\
\sin \varepsilon &=& \sqrt{\sin^2 K_2 \cos^2 K_1 + \sin^2 K_1} \nonumber
\end{eqnarray}
which allows us to write the Floquet operators as 

\begin{equation}
\hat{U}_{T,i}^\mathrm{MF} = \cos \varepsilon - i \sin \varepsilon \left ( n_{ix} \hat{\sigma}_x + n_{iy} \hat{\sigma}_y \right )
\label{eq:UTGen}
\end{equation}
where the vector components are

\begin{eqnarray}
n_{1x} &=& \frac{\sin K_1 \cos K_2}{\sin \varepsilon}, \hspace{20pt} n_{1y} = \frac{\sin K_2}{\sin \varepsilon} \\
n_{2x} &=& \frac{\sin K_1}{\sin \varepsilon}, \hspace{50pt} n_{2y} = \frac{\sin K_2 \cos K_1}{\sin \varepsilon} \, .
\end{eqnarray}
It is clear from Eq.\ \eqref{eq:UTGen} that the effective Hamiltonian mentioned in the main text takes the form  $\hat{H}_{\mathrm{eff},i}^\mathrm{MF} = \varepsilon \boldsymbol{n_i} \cdot \hat{\boldsymbol{\sigma}}$.  We note that the derivation of the winding vector components is the same as the one for the spin-1/2 QDKR in Ref.\ \cite{zhou18} with the only difference being that here we have $K_1 = \kappa_1 \sqrt{1-(m/j)^2} \cos \phi$ and $K_2 = \kappa_2 \sqrt{1-(m/j)^2} \sin \phi$ and they have $m = 0$.

\section{Short and long time averages of the chiral displacement \label{app:LTA}}

\begin{figure}[t]
\centering
\includegraphics[width=\columnwidth]{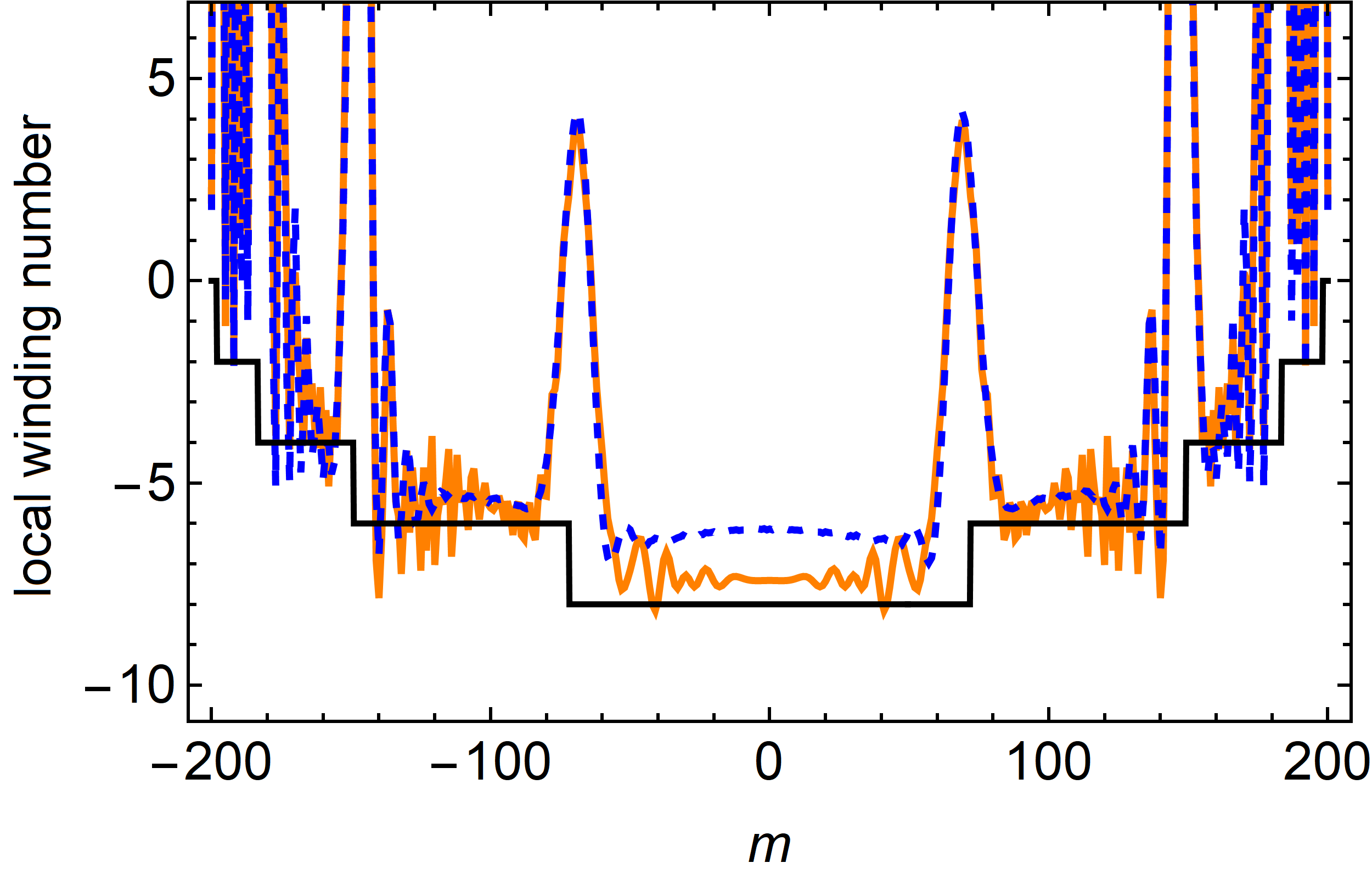}
\caption{Comparison of short time and long time averages of the chiral displacement.  The solid orange and dashed blue curves show time averages of the chiral displacement using Eq.\ \eqref{eq:TA} for $\mathcal{N} = 10$ and $\mathcal{N} = 1000$, respectively.  The long time average has less agreement with the mean field prediction (solid black) than the short time average because important terms get 'washed-out' in the averaging process.  The parameter values are $\kappa_1 = 7.5 \pi$, $\kappa_2 = 0.5 \pi$ and $j = 200$.}
\label{fig:TAWN}
\end{figure}

We find that the short time average rather than the long time average of the chiral displacement in Eq.\ \eqref{eq:CD} captures the winding number.  Figure \ref{fig:TAWN} explicitly shows the difference between short and long time average calculations of $w_\pi$ for $\kappa_1 = 7.5 \pi$ and $\kappa_2 = 0.5 \pi$.  The solid orange and dashed blue data is calculated with $\mathcal{N} = 10$ and $\mathcal{N}= 1000$, respectively, and the solid black curve is the mean field prediction.  The long time average result has less fluctuations than the short time average, but has a larger disagreement with the mean field result over the range of states.  This is because the long time average does not keep terms like $\langle \varepsilon_k \vert \hat{J}_z \hat{\sigma}_z \vert \varepsilon_j \rangle$, where $j \neq k$, which are necessary to accurately predict the winding number from the chiral displacement.  However, the long time average still exhibits the sudden jumps at the boundaries, so it can still be used to locate the boundaries.


\begin{thebibliography}{8}

\bibitem{sacha15}{K. Sacha, Anderson localization and Mott insulator phase in the time domain, Sci. Rep. \textbf{5}, 10787 (2015).}

\bibitem{giergiel17}{K. Giergiel and K. Sacha, Anderson localization of a Rydberg electron along a classical orbit, Phys. Rev. A \textbf{95}, 063402 (2017).}

\bibitem{guo13}{L. Guo, M. Marthaler, and G. Sch\"{o}n, Phase Space Crystals: A New Way to Create a Quasisenergy Band Structure, Phys. Rev. Lett. \textbf{111}, 205303 (2013).}

\bibitem{guo16}{L. Guo and M. Marthaler, Synthesizing lattice structures in phase space, New J. Phys. \textbf{18}, 023006 (2016).}

\bibitem{goldman14a}{N. Goldman, G. Juzeli$\bar{\mathrm{u}}$nas, P. \"{O}hberg, and I. B. Spielman, Light-induced gauge fields for ultracold atoms, Rep. Prog. Phys. \textbf{77}, 126401 (2014).}

\bibitem{goldman14b}{N. Goldman and J. Dalibard, Periodically driven quantum systems: effective Hamiltonians and engineered gauge fields, Phys. Rev. X \textbf{4} 031027 (2014).}

\bibitem{eckardt15}{A. Eckardt and E. Anisimovas, High-frequency approximation for periodically driven quantum systems from a Floquet space perspective, New J. Phys. \textbf{17}, 093039 (2015).}

\bibitem{bukov15}{M. Bukov, L. D'Alessio, and A. Polkovnikov, Universal high-frequency behavior of periodically driven systems: from dynamical stabilization to Floquet engineering, Adv. Phys. \textbf{64}, 139-226 (2015).}

\bibitem{lindner11}{N. H. Lindner, G. Refael, and V. Galitski, Floquet topological insulator in semiconductor quantum wells, Nature Phys. \textbf{7}, 490 (2011).}


\bibitem{jaksch05}{D. Jaksch and P. Zoller, The cold atom Hubbard toolbox, Ann. Phys. \textbf{315}, 52 (2005).}

\bibitem{dalibard11}{J. Dalibard, F. Gerbier, G. Juzeli$\bar{\mathrm{u}}$nas, and P \"{O}hberg, Colloquium: Artificial Gauge Potentials for Neutral Atoms, Rev. Mod. Phys. \textbf{83}, 1523 (2011).}

\bibitem{creffield16}{C. E. Creffield, G. Pieplow, F. Sols, and N. Goldman, Realization of uniform synthetic magnetic fields by periodically shaking an optical square lattice, New J. Phys. \textbf{18}, 093013 (2016).}

\bibitem{stuhl15}{B. K. Stuhl, H.-I. Lu, L. M. Aycock, D. Genkina, and I. B. Spielman, Visualizing edge states with an atomic Bose gas in the quantum Hall regime, Science \textbf{349}, 1514-1518 (2015).}

\bibitem{aidelsburger13}{M. Aidelsburger, M. Atala, M. Lohse, J. T. Berreiro, B. Paredes, and I Bloch, Realization of the Hofstadter Hamiltonian with ultracold atoms in optical lattices, Phys. Rev. Lett. \textbf{111}, 185301 (2013).}

\bibitem{miyake13}{H. Miyake, G. A. Siviloglou, C. J. Kennedy, W. C. Burton, and W. Ketterle, Realizing the Harper Hamiltonian with laser-assisted tunneling in optical lattices, Phys. Rev. Lett. \textbf{111}, 185302 (2013).}


\bibitem{leboeuf90}{R. Leboeuf, J. Kurchan, M. Feingold, and D. P. Arovas, Phase-space localization: Topological aspects of quantum chaos, Phys. Rev. Lett. \textbf{65}, 3076 (1990).}

\bibitem{jangjan22}{M. Jangjan, L. E. F. Foa Torres, and M. V. Hosseini, Floquet topological phase transitions in a periodically quenched dimer, Phys. Rev. B \textbf{106}, 224306 (2022).}

\bibitem{cedzich18}{C. Cedzich, T. Geib, C. Stahl, L. Vel\'{a}zquez, A. H. Werner, and R. F. Werner, Complete homotopy invariants for translation invariant symmetric quanstum walks on a chain, Quantum \textbf{2}, 95 (2018).}

\bibitem{cedzich21}{C. Cedzich, T. Geib, A. H. Werner, and R. F. Werner, Chiral Floquet Systems and Quantum Walks at Half-Period, Ann. Henri Poincar\'{e} \textbf{22}, 375-413 (2021).} 

\bibitem{wang08}{J. Wang and J. B. Gong, Proposal of a cold-atom realization of quantum maps with Hofstadter's butterfly spectrum, Phys. Rev. A \textbf{77}, 031405(R) (2008); J. Wang and J. Gong, Butterfly Floquet Spectrum in Driven SU(2) Systems, Phys. Rev. Lett. \textbf{102}, 244102 (2009).}

\bibitem{wang09}{J. Wang, A. S. Mouritzen, and J. Gong, Quantum control of ultra-cold atoms: uncovering a novel connection between two paradigms of quantum nonlinear dynamics, J. Mod. Optics \textbf{56}, 722 (2009).}

\bibitem{ho12}{D. Y. H. Ho and J. Gong, Quntized Transport in Momentum Space, Phys. Rev. Lett. \textbf{109}, 010601 (2012).}






\bibitem{zhou18}{L. Zhou and J. Gong, Floquet topological phases in a spin-1/2 double kicked rotor, Phys. Rev. A \textbf{97}, 063603 (2018).}

\bibitem{bolik22}{N. Bolik, C. Groiseau, J. H. Clark, G. S. Summy, Y. Liu, and S. Wimberger, Detecting topological phase transitions in a double kicked quantum rotor, Phys. Rev. A \textbf{106}, 043318 (2022).}

\bibitem{aharonov93}{Y. Aharonov, L. Davidovich, and N. Zagury, Quantum random walks, Phys. Rev. A \textbf{48}, 1687 (1993).}

\bibitem{farhi98}{E. Farhi and S. Gutmann, Quantum computation and decision trees, Phys. Rev. A \textbf{58}, 915 (1998).}

\bibitem{shenvi03}{N. Shenvi, J. Kempe, and K. B. Whaley, Quantum Random Walk Search Algorithm, Phys. Rev. A \textbf{67}, 052307 (2003).}

\bibitem{childs09}{A. M. Childs, Universal Computation by Quantum Walks, Phys. Rev. Lett. \textbf{102}, 180501 (2009).}

\bibitem{kitagawa10}{T. Kitagawa, M. S. Rudner, E. Berg, and E. Demler, Exploring topological phases with quantum walks, Phys. Rev. A \textbf{82}, 033429 (2010); T. Kitagawa, Topological phenomena in quantum walks: elementary introduction to the physics of topological phases, Quantum Inf. Process. \textbf{11}, 1107 (2012).}

\bibitem{groh16}{T. Groh, S. Brakhane, W. Alt, D. Meschede, J. K. Asb\'{o}th, and A. Alberti, Robustness of topologically protected edge states in quantum walk experiments with neutral atoms, Phys. Rev. A \textbf{94}, 013620 (2016).}

\bibitem{sajid19}{M. Sajid, J. K. Asb\'{o}the, D. Meschede, R. F. Werner, and A. Alberti, Creating anomalous Floquet Chern insulator with magnetic quantum walks, Phys. Rev. B \textbf{99}, 214303 (2019).}

\bibitem{cardano17}{F. Cardano, A. D'Errico, A. Dauphin, M. Maffei, B. Piccirillo, C. de Lisio, G. D. Filippis, V. Cataudella, E. Santamato, L. Marrucci, M. Lewenstein, and P. Massignan, Detection of Zak phases and topological invariants in a chiral quantum walk of twisted photons, Nat. Commun. \textbf{8}, 15516 (2017).}

\bibitem{xu18}{X.-Y. Xu, Q.-Q. Wang, W.-W Pan, K. Sun, J.-S. Xu, G. Chen, J.-S. Tang, M. Gong, Y.-J. Han, C.-F. Li, and G.-C. Guo, Measuring the Winding Number in a Large-Scale Chiral Quantum Walk, Phys. Rev. Lett. \textbf{120}, 260501 (2018).}

\bibitem{kitagawa12}{T. Kitagawa, M. A. Broome, A. Fedrizzi, M. S. Rudner, E. Berg, I. Kassal, A. Aspuru-Guzik, E. Demler, and A. G. White, Observation of topologically protected bound states in photonic quantum walks, Nat. Commun. \textbf{3}, 882 (2012).}

\bibitem{wang18}{B. Wang, T. Chen, and X. Zhang, Experimental Observation of Topologically Protected Bound States with Vanishing Chern Numbers in a Two-Dimensional Quantum Walk, Phys. Rev. Lett. \textbf{121} 100501 (2018).}

\bibitem{chen18}{C. Chen, X. Ding, J. Qin, Y. He, Y.-H. Luo, M.-C. Chen, C. Liu, X.-L. Wang, W.-J. Zhang, H. Li, L.-X. You, Z. Wang, D.-W. Wang, B. C. Sanders, C.-Y. Lu, and J.-W. Pan, Observation of Topologically Protected Edge States in a Photonic Two-Dimensional Quantum Walk, Phys. Rev. Lett. \textbf{121}, 100502 (2018).}

\bibitem{haake88}{F. Haake and D. L. Shepelyansky, The Kicked Rotator as a Limit of the Kicked Top, Europhys. Lett. \textbf{5} (8), 671 (1988).}

\bibitem{nitsche19}{T. Nitsche, T. Geib, C. Stahl, L. Lorz, C. Cedzich, S. Barkhofen, R. F. Werner, and C. Silberhorn, Eigenvalue measurement of topologically protected edge states in split-step quantum walks, New J. Phys. \textbf{21}, 043031 (2019).}

\bibitem{biederharn81}{L. G. Biederharn and J. C. Louck, \textit{Angular Momentum in Quantum Physics: Theory and Applications} (Addison-Wesley, Reading, MA, 1981).}

\bibitem{zibold10}{T. Zibold, E. Nicklas, C. Gross, and M. K. Oberthaler, Classical Bifurcation at the Transition from Rabi to Josephson Dynamics, Phys. Rev. Lett. \textbf{105}, 204101 (2010).}

\bibitem{gerving12}{C. S. Gerving, T. M. Hoang, B. J. Land, M. Anquez, C. D. Hamley, and M. S. Chapman, Non-equilibrium dynamics of an unstable quantum pendulum explored in a spin-1 Bose-Einstein condensate, Nat. Commun. \textbf{3}, 1169 (2012).}

\bibitem{bao20}{H. Bao, J. Duan, S. Jin, X. Lu, P. Li, W. Qu, M. Wang, I. Novikova, E. E. Mikhailov, K.-F. Zhao, K. M{\o}lmer, H. Shen, and Y. Xiao, Spin squeezing of $10^{11}$ atoms by prediction and retrodiction measurements, Nature \textbf{581}, 159 (2020).}

\bibitem{mulansky11}{F. Mulansky, J. Mumford, and D. H. J. O'Dell, Impurity in a Bose-Einstein condensate in a double well, Phys. Rev. A \textbf{84}, 063602 (2011).}

\bibitem{mumford14}{J. Mumford, J. Larson, and D. H. J. O'Dell, Impurity in a bosonic Josephson junction: Swallowtail loops, chaos, self-trapping, and Dicke model, Phys. Rev. A \textbf{89}, 023620 (2014).}

\bibitem{talukdar10}{I. Talukdar, R. Shrestha, and G. S. Summy, Sub-Fourier Characteristics of a $\delta$-kicked-rotor Resonance, Phys. Rev. Lett. \textbf{105}, 054103 (2010).}

\bibitem{kanem07}{J. F. Kanem, S. Maneshi, M. Partlow, M. Spanner, and A. M. Steinberg, Observation of High-Order Quantum Resonances in the Kicked Rotor, Phys. Rev. Lett. \textbf{98}, 083004 (2007).}

\bibitem{ullah11}{A. Ullah and M. D. Hoogerland, Experimental observation of Loschmidt time reversal of a quantum chaotic system, Phys. Rev. E \textbf{83}, 046218 (2011).}


\bibitem{huang21}{L.-G. Huang, F. Chen, X. Li, Y. Li, R. L\"{u}, and Y.-C. Liu, Dynamic synthesis of Heisenberg-limited spin squeezing, npj Quantum Inform. \textbf{7}, 168 (2021).}

\bibitem{will11}{S. Will, T. Best, S. Braun, U. Schneider, and I. Bloch, Coherent Interaction of a Single Fermion with a Small Bosonic Field, Phys. Rev. Lett. \textbf{106}, 115305 (2011).}

\bibitem{duan22}{L. Duan, Quantum walk on the Bloch sphere, Phys. Rev. A \textbf{105}, 042215 (2022).}

\bibitem{roy17}{R. Roy and F. Harper, Periodic table for Floquet topological insulators, Phys. Rev. B \textbf{96}, 155118 (2017).} 

\bibitem{asboth12}{J. K. Asb\'{o}th, Symmetries, topological phases, and bound states in the one-dimensional quantum walk, Phys. Rev. B \textbf{86}, 195414 (2012); J. K. Asb\'{o}th and H. Obuse, Bulk-boundary correspondence for chiral symmetric quantum walks, Phys. Rev. B \textbf{88} 121406(R) (2013).} 

\bibitem{siemens23}{A. Siemens and P. Schmelcher, Geometry induced domain-walls of dipole lattices on curved structures, arXiv:2302.13728.}

\bibitem{meier18}{E. J. Meier, F. A. An, A. Dauphin, M. Maffei, P. Massignan, T. L. Hughes, and B. Gadway, Observation of the topological Anderson insulator in disordered atomic wires, Science \textbf{362}, 929 (2018).}

\bibitem{song14}{J. Song and E. Prodan, AIII and BDI topological systems at strong disorder, Phys. Rev. B \textbf{89}, 224203 (2014).}

\bibitem{shem14}{I. Mondragon-Shem, T. L. Hughes, J. Song, and E. Prodan, Topological criticality in the chiral symmetric AIII class at strong disorder, Phys. Rev. Lett. \textbf{113}, 046802 (2014).}

\bibitem{maffei18}{M Maffei, A. Dauphin, F. Cardano, M. Lewenstein, and P. Massignan, Topological characterization of chiral models through their long time dynamics, New. J. Phys. \textbf{20}, 013023 (2018).}

\bibitem{vlastakis15}{B. Vlastakis, A. Petrenko, N. Ofek, L. Sun, Z. Leghtas, K. Sliwa, Y. Liu, M. Hatridge, J. Blumoff, L. Frunzio, M. Mirrahimi, L. Jiang, M. H. Devoret, and R. J. Schoelkopf, Characterizing entanglement of an artificial atom and cavity state with Bell's inequality, Nat. Commun. \textbf{6}, 8970 (2015).}

\bibitem{sone22}{K. Sone, Y. Ashida, and T. Sagawa, Topological synchronization of coupled nonlinear oscillators, Phys. Rev. Research \textbf{4}, 023211 (2022).}






































































































\end{thebibliography}
\end{document}